\documentclass[final,5p,times,twocolumn]{elsarticle}

\usepackage{amssymb}

\usepackage{graphicx}

\journal{Physics Letters B}

\begin{document}
\begin{frontmatter}

\title{Charge radii and structural evolution in Sr, Zr, and Mo 
isotopes}

\author[csic]{R. Rodr\'{\i}guez-Guzm\'an} 
\author[csic]{P. Sarriguren}
\author[uam]{L.M. Robledo}
\author[uam]{S. Perez-Martin\fnref{ciemat}}

\address[csic]{Instituto de Estructura de la Materia, CSIC, Serrano
123, E-28006 Madrid, Spain}

\address[uam]{Departamento  de F\'{\i}sica Te\'orica, M\'odulo 15,
Universidad Aut\'onoma de Madrid, 28049-Madrid, Spain.}

\fntext[ciemat] {Currently at CIEMAT Nuclear Fission Division,
 Av. Complutense 22, Edif. 17, 28040 Madrid (SPAIN)}



\begin{abstract} 
The evolution of the ground-state nuclear shapes in neutron-rich 
Sr, Zr, and Mo isotopes, including both even-even and odd-$A$ 
nuclei, is studied within a self-consistent mean-field approximation 
based on the D1S Gogny interaction.
Neutron separation energies and charge radii are calculated and
compared with available data. A correlation between a shape 
transition and a discontinuity in those observables is found
microscopically. While in Sr and Zr isotopes the steep behavior 
observed in the isotopic dependence of the charge radii is a 
consequence of a sharp prolate-oblate transition, the smooth 
behavior found in Mo isotopes has its origin in an emergent 
region of triaxiality.

\end{abstract}

\end{frontmatter}




\section{Introduction}

The study of the properties of unstable nuclei both theoretically 
and experimentally is nowadays one of the most active and fruitful 
research lines in nuclear physics.
Nuclear systems with very unusual $N/Z$ ratios are proper candidates
to get insight into the nuclear interaction and the impact that the
associated dynamics might have in low-energy observables like the
ground-state deformation and derived quantities like moments of
inertia or vibrational excitation energies \cite{review}. 
The understanding of the properties of those nuclei also has 
important consequences in the understanding of other physical 
processes like stellar nucleosynthesis of heavy elements.
In particular, the neutron-rich Sr, Zr, and Mo isotopes with mass 
numbers $A=100-110$ are of special interest for various reasons. 
One of them is the role that these isotopes play in the 
nucleosynthesis of heavy nuclides in the astrophysical r process. 
Their masses and decay properties are an essential input 
to model the path, the isotopic abundances and the time scale of 
the r process in a reliable way \cite{cowan}. In addition, from the 
nuclear structure point of view, this region is characterized 
by a strong competition between various shapes, giving rise to shape 
instabilities that lead to coexisting nuclear shapes, as well as to 
sudden shape transitions in isotopic chains \cite{wood92}.

It has been shown that the ground states of Sr, Zr and Mo isotopes 
with $N$ ranging from the magic number $N=50$ up to $N\sim 60$ are 
weakly deformed, but they undergo a shape transition from nearly 
spherical to well deformed prolate (or oblate) deformations as $N= 60$ 
is approached and crossed. 
Evidence for such an abrupt shape change includes $2^+$ lifetime
measurements \cite{mach,goodin} as well as quadrupole moments for 
rotational bands \cite{urban}. Signatures of triaxiality in Mo 
isotopes from spectroscopic studies of high-spin states have also 
been identified \cite{hua}.
Heavier Sr and Zr ($A\sim 110$) isotopes display an axially symmetric 
well deformed shape.  Above this region, it has been suggested that 
the $N=82$ shell closure might be quenched far from stability. 
This quenching has been predicted by different models \cite{doba96}, 
but still weak experimental evidence has been found.

Masses (nuclear binding energies) and charge radii are considered 
among the most sensitive observables to explore the structural 
evolution in nuclei far from the valley of stability, where other 
types of spectroscopic experiments are prohibitive because of the 
low production rates and short lifetimes of these unstable isotopes.
Discontinuities in the systematics of particular mass regions
suggest a change in the structure of the ground states among 
the configurations competing for the lowest energy.
Drastic changes in the mean square (ms) radius of the charge 
distributions in regions of transitional nuclei may also be 
indicators of structural changes related to the nuclear 
deformation. 

\section{Theoretical framework}

In this work we study this phenomenology in the framework of the 
self-consistent Hartree-Fock-Bogoliubov (HFB) approximation 
based on the finite range and density dependent Gogny interaction  
\cite{gogny} with the parametrization D1S \cite{d1s}.
The HFB wave functions are expanded in a harmonic oscillator 
basis containing  enough number of shells to achieve convergence 
of the results for all the nuclei studied.

The quadrupole deformations at equilibrium are obtained self-consistently 
by a minimization procedure. Nevertheless, constrained calculations have 
also been performed to generate potential energy curves (PEC) or potential 
energy surfaces (PES) to study the evolution  with the number of nucleons
of the various shapes associated to the different minima displayed by
the PECs or PESs. This has been accomplished preserving axial symmetry 
first, and allowing afterward for triaxiality in the most relevant isotopes 
where shape transitions take place. Hence, we have constrained the mean value 
of the quadrupole operators $\hat Q_{20}$ and $\hat Q_{22}$ (in the axial case 
only the former is constrained),

\begin{equation}
\label{Q20}
Q_{20} = \frac{1}{2} \langle \Phi_{\rm HFB} | 2z^{2} - 
x^{2} - y^{2}  | \Phi_{\rm HFB} \rangle \, ,
\end{equation}

\begin{equation}
\label{Q22}
Q_{22} = \frac{\sqrt{3}}{2} \langle \Phi_{\rm HFB} |  x^{2} - y^{2}  | 
\Phi_{\rm HFB} \rangle \, .
\end{equation}
 In the triaxial figures we plot  $Q-\gamma$ planes, where 
\begin{equation}
\label{Q0}
Q = \sqrt{  Q_{20}^{2}+ Q_{22}^{2}}
\end{equation}
and $\gamma$ is the angle defined as $\tan \gamma =Q_{22}/Q_{20}$
\cite{ours}. 
With this definition an axially symmetric prolate mass distribution 
has $\gamma=0 ^\circ$, whereas the corresponding oblate one has 
$\gamma=60 ^\circ$. The quadrupole deformation parameter $\beta$ is 
defined in terms of the mass quadrupole moment $Q_{20}$ and ms radius 
$\langle r^2 \rangle$,
\begin{equation}
\beta = \sqrt{\frac{\pi}{5}}\frac{Q_{20}}{A\langle r^2 \rangle}\, .
\label{beta_quadru}
\end{equation}

For the description of odd-$A$ nuclei, we have extended the HFB 
formalism using blocking techniques. The blocked HFB wave function of 
the odd-A system is in general given by
\begin{equation}
\label{odd}
| \Phi_{\rm HFB} \rangle _{\alpha} 
= \beta^+_{\alpha} \, | \Phi_{\rm HFB} \rangle \, ,
\end{equation}
where $| \Phi_{\rm HFB} \rangle $ is a reference even-even HFB vacuum 
($\beta_{\alpha}  | \Phi_{\rm HFB} \rangle =0$) and $\beta^+_{\alpha}$ is a 
quasiparticle creation operator. The index $\alpha$ stands for the 
quasiparticle quantum numbers characterizing the blocked state 
(angular momentum projection $K$ and parity in the case of axial 
symmetry). The ground state of the odd nucleus is determined by 
finding the blocked state that minimizes the total energy.
In the present study we use the equal filling approximation (EFA), a 
prescription widely used in mean-field calculations to treat the 
dynamics of odd nuclei in a time-reversal invariant way. In this 
approximation the unpaired nucleon is treated in an equal footing 
with its time-reversed state by sitting half a nucleon in a given 
orbital and the other half in the time-reversed partner. This 
procedure has been recently justified microscopically, showing 
that the EFA can be described in terms of a mixed state density
operator and the equations to be solved are a direct consequence 
of the variational principle over the energy of such mixed state 
\cite{perez}. It has also been shown that the EFA and the exact 
blocking procedure are both strictly equivalent when the time-odd 
fields of the energy density functional are neglected \cite{schunck}. 
Thus, EFA is sufficiently precise for most practical applications.

In this work, we present results for one-neutron ($S_n$) and 
two-neutron ($S_{2n}$)  separation energies,
which can be easily calculated from the binding energies $BE$,

\begin{eqnarray}
S_n (Z,N) &=& -BE(Z,N)+BE(Z,N-1), \nonumber \\ 
S_{2n}(Z,N) &=& -BE(Z,N)+BE(Z,N-2)\ .
\end{eqnarray}

Charge radii and their differences are crucial quantities to study 
the systematics of the nuclear-shape changes as they can be measured 
with remarkable precision using laser spectroscopic techniques. They 
are obtained theoretically by folding the proton distribution of the 
nucleus with the the finite size of the protons and the neutrons. 
The ms radius of the charge distribution in a nucleus can be expressed 
as
\begin{equation}
\langle r^2_c \rangle = \langle r^2_p \rangle _Z+
\langle r^2_c \rangle _p +(N/Z)
\langle r^2_c \rangle _n + r^2_{CM} 
\, , \label{rch}
\end{equation}
where $ \langle r^2_p \rangle _Z$ is the ms radius of the 
point proton distribution in the nucleus

\begin{equation}
 \langle r_p^2 \rangle _Z = \frac{ \int r^2\rho_p({\vec r})d{\vec r} }
{\int \rho_p({\vec r})d{\vec r}} \, , \label{r2pn}
\end{equation}
$ \langle r^2_c \rangle _p=0.80$ fm$^2$ and 
$ \langle r^2_c \rangle _n=-0.12$ fm$^2$ are the ms radii 
of the charge distributions in a proton and a neutron, respectively. 
$r^2_{CM}$ is a small correction due to the center of mass motion, 
which is evaluated according to Ref. \cite{negele}. It is worth 
noticing that the most important correction to the point proton ms 
nuclear radius, coming from the proton charge distribution 
$ \langle r^2_c \rangle _p$, vanishes when isotopic differences are 
considered, since it does not involve any dependence on $N$.

The variations of the charge radii in isotopic chains are related
to deformation effects. Other structure effects like pairing
or spin-orbit couplings have been considered \cite{tajima,reinhard},
but in this work they are fixed by the Gogny interaction. 
For an axially symmetric static quadrupole deformation $\beta$ 
the increase of the charge radius with respect to the spherical 
value is given to first order by

\begin{equation}
\langle r^2 \rangle = \langle r^2 \rangle _{\rm sph} \left(
1+\frac{5}{4\pi} \beta^2 \right) \, ,
\end{equation}
where usually $\langle r^2 \rangle _{\rm sph}$ is taken from the 
droplet model. Then, the measured radii have been used to estimate 
the changes in quadrupole deformation \cite{charlwood}.
On the contrary, in this work we analyze the effect of the quadrupole 
deformation on the charge radii from a microscopic self-consistent 
approach.

\begin{figure}[ht]
\centering
\includegraphics[width=80mm]{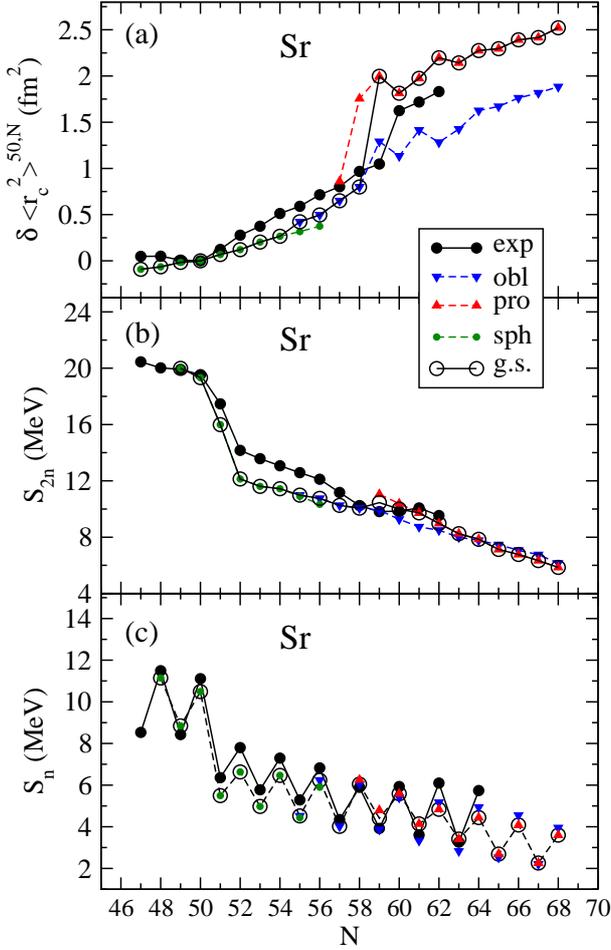}
\caption{Calculated $\delta \langle r^2_c \rangle$ (a), $S_{2n}$ (b), and
$S_n$ (c) in Sr isotopes compared to experimental data from Ref. 
\cite{hager06} for masses and from Ref. \cite{buchinger} for radii.
Results for prolate, oblate, and spherical minima are displayed with
different symbols (see legend). Open circles correspond to ground-state
results. }
\label{fig_sr_r2}
\end{figure}

\begin{figure}[ht]
\centering
\includegraphics[width=80mm]{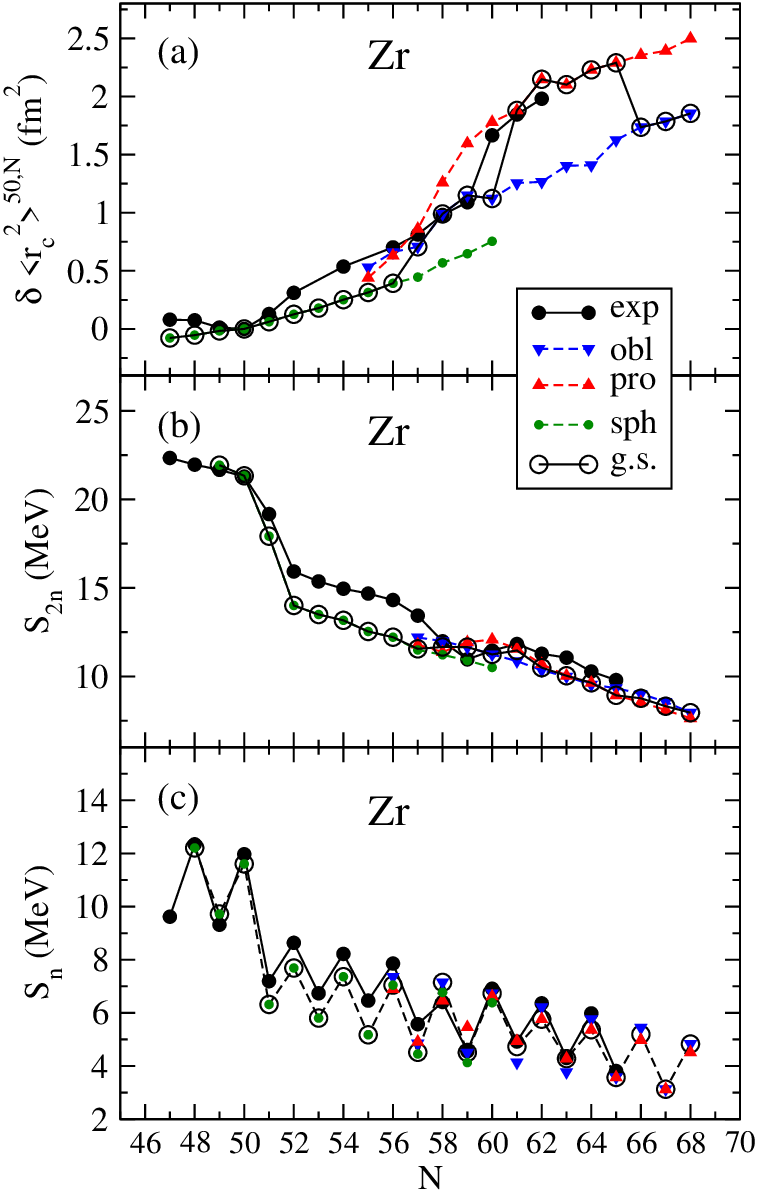}
\caption{Same as in Fig. \ref{fig_sr_r2}, but for Zr isotopes. Experimental 
data for masses and radii are from Refs. \cite{hager06} and \cite{campbell}, 
respectively.}
\label{fig_zr_r2}
\end{figure}

\begin{figure}[ht]
\centering
\includegraphics[width=80mm]{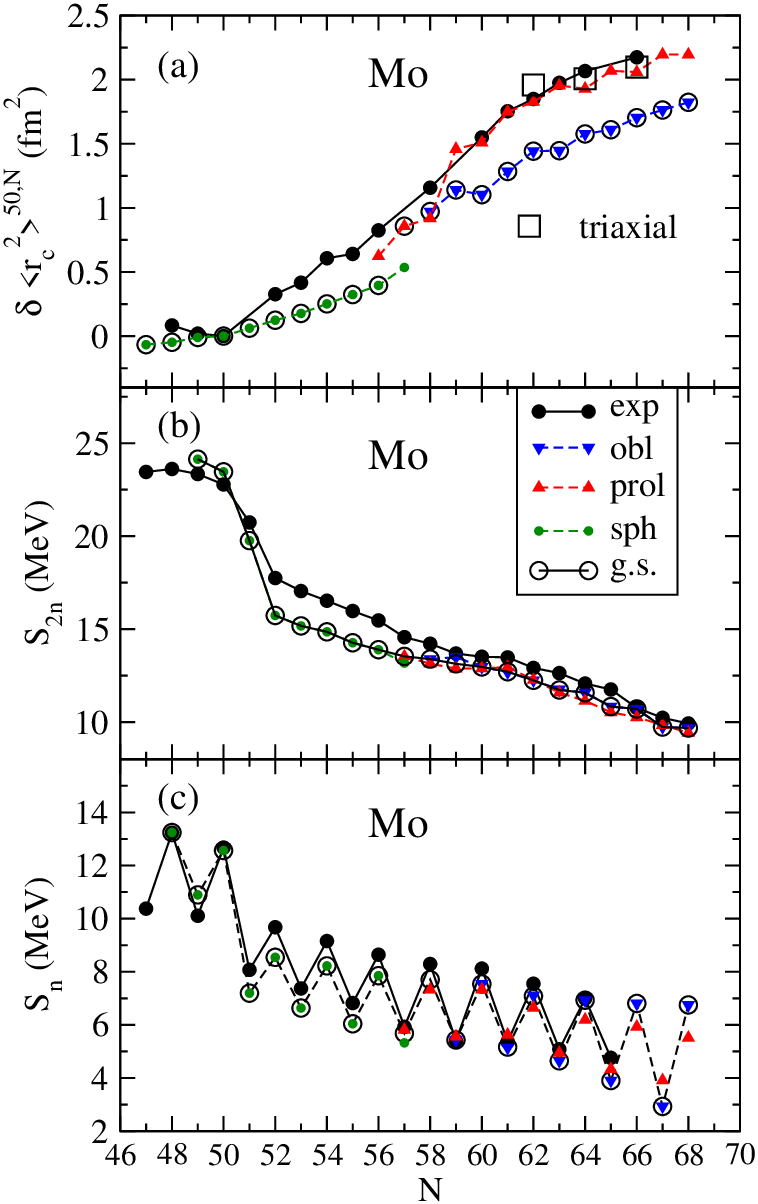}
\caption{Same as in Fig. \ref{fig_sr_r2}, but for Mo isotopes. Experimental 
data for masses and radii are from Refs. \cite{hager06} and \cite{charlwood}, 
respectively.}
\label{fig_mo_r2}
\end{figure}

\section{Results}

The PECs of the neutron-rich isotopes of Sr, Zr, and Mo from $N=50$ to
$N=68$ have been calculated with the Gogny-D1S interaction. Our results
fully agree with those of Ref. \cite{webpage} that were obtained in the 
same framework. The main features of the shape evolution in the axial
case can be summarized as follows:
The isotopes with $N=50-54$ show a sharp PEC around the spherical minimum 
that becomes rather shallow at $N=56-58$.
Isotopes with $N=60$ are already 
deformed with oblate and prolate minima very close in  energy. In the 
case of Sr isotopes the ground state is prolate, for Zr isotopes both 
oblate and prolate minima are found at about the same energy, while 
for Mo isotopes the ground state is oblate. Beyond $N=60$ the shapes 
become stable and for heavier isotopes we obtain basically 
similar results to $N=60$.
It is also worth mentioning the incipient emergence of
a spherical solution in the heavier isotopes.

The shape change at $N\sim 60$ has been predicted to a different 
extent by various theoretical models. Global calculations within the 
finite-range droplet model \cite{moller95} with single-particle
states obtained from folded Yukawa predict prolate quadrupole
deformations increasing smoothly from $N=50$ up to $N=56$, and then
jumping suddenly to large deformations between $N=58$ and $N=60$. 
At $N=62$ the shapes stabilize until a transition to oblate shapes 
is predicted at $N=74$ in Sr and Zr, and at $N=72$ in Mo. Triaxial 
calculations within this approach were carried out in Ref. \cite{moller08},
showing that only Mo isotopes have a tendency to triaxiality in this 
mass region. 

The approach followed in Ref. \cite{skalski}, where PESs were studied 
within a finite-range liquid drop model modified by shell corrections 
taken from deformed Woods-Saxon potentials, suggests an oblate-prolate 
shape coexistence in Sr and Zr isotopes from $N=60-70$ with prolate 
ground states. Mo isotopes display in this case a soft behavior that 
develops triaxiality at $N=72,74$.
Other calculations including rotational states in terms of the total 
Routhian surface, using non-axial Wood-Saxon potentials \cite{xu02}, 
predict two coexisting prolate and oblate minima for $^{106-116}$Zr 
isotopes, where the prolate ground state becomes oblate beyond $^{110}$Zr. 
The same calculations predict a $\gamma$-soft triaxial minimum for 
$^{108}$Mo. 

\begin{figure}[ht]
\centering
\includegraphics[width=80mm]{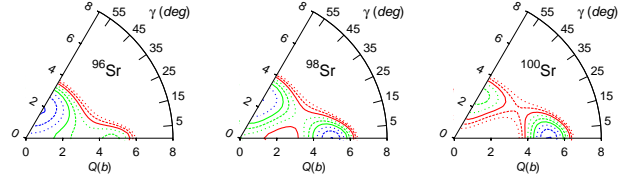}
\caption{$Q-\gamma$ planes for $^{96,98,100}$Sr isotopes with the 
Gogny-D1S interaction. The contour lines extend from the minimum 
up to 2 MeV higher in steps of 0.25 MeV.}
\label{fig_sr_tri}
\end{figure}

\begin{figure}[ht]
\centering
\includegraphics[width=80mm]{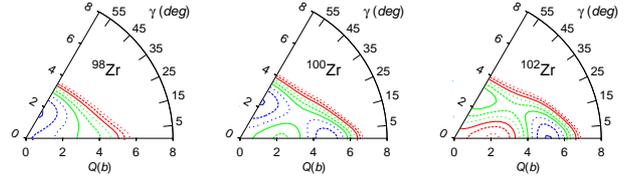}%
\caption{Same as in Fig. \ref{fig_sr_tri}, but for $^{98,100,102}$Zr isotopes.}
\label{fig_zr_tri}
\end{figure}

\begin{figure}[ht]
\centering
\includegraphics[width=80mm]{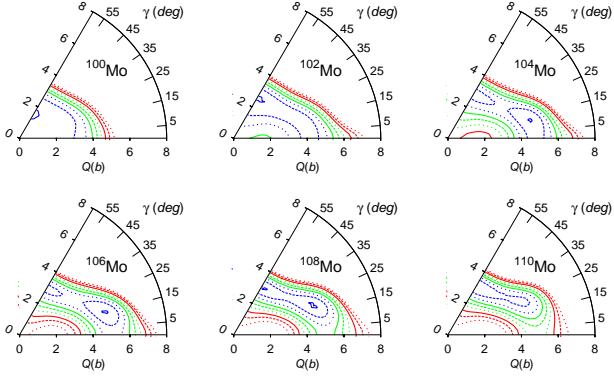}
\caption{Same as in Fig. \ref{fig_sr_tri}, but for $^{100-110}$Mo isotopes.}
\label{fig_mo_tri}
\end{figure}

\begin{figure}[ht]
\centering
\includegraphics[width=75mm]{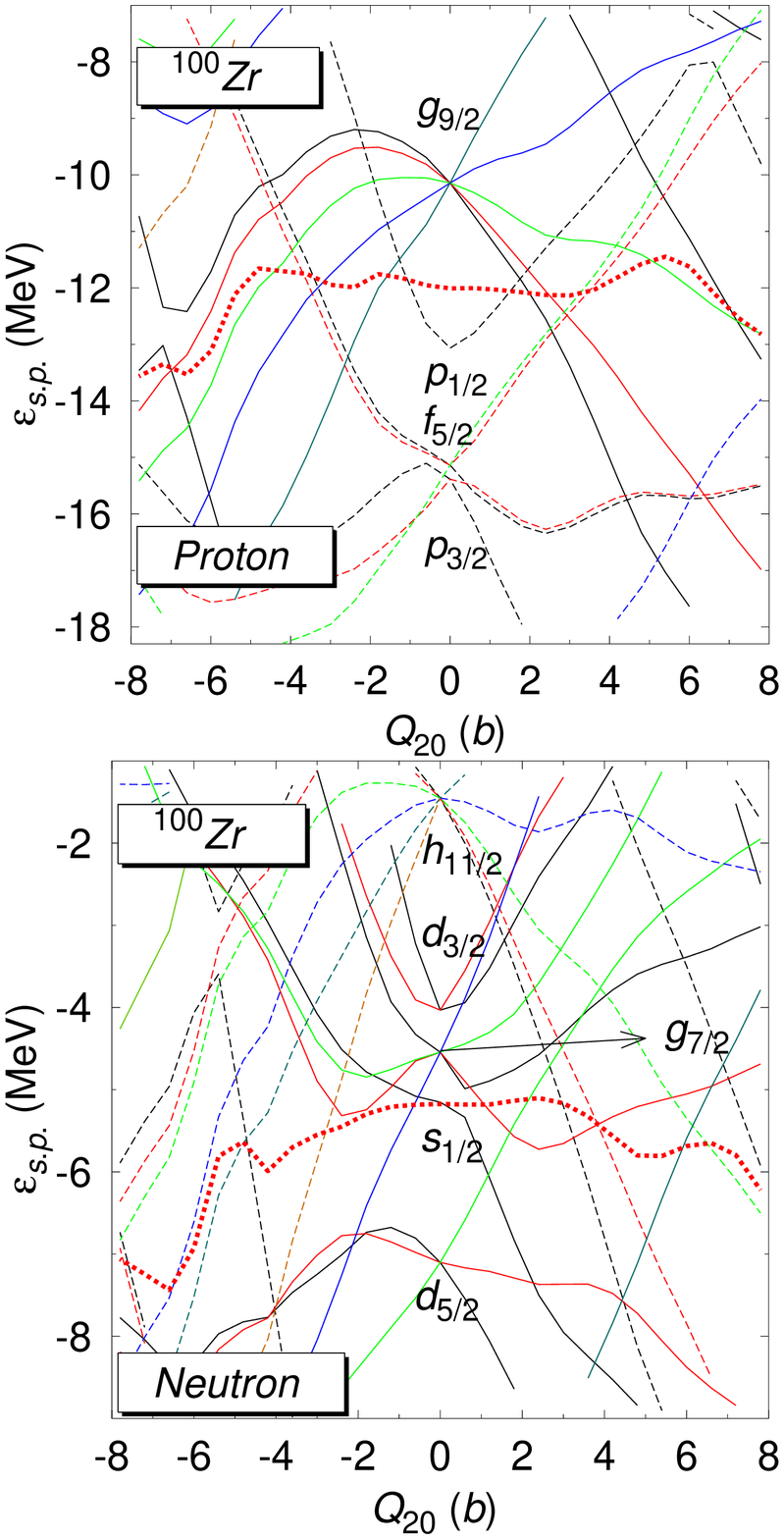}
\caption{Proton and neutron SPEs for $^{100}$Zr ($Z=40,N=60$) as a function
of the axial quadrupole moment $Q_{20}$ (b). The Fermi level is plotted
by a thick dotted line.}
\label{fig_zr_sp}
\end{figure}

Relativistic mean-field calculations \cite{lala} also show  increasing
prolate deformations from $N=50$ up to $N=60$. In the region between
$N=60$ and $N=70$ the deformations are stable for both  Sr and Zr, 
whereas they change to oblate at $N=64$ in Mo.
Triaxial PESs calculated from Skyrme HF+BCS were already published 
in Ref. \cite{bonche85}, where the shape evolution along Kr, Sr, Zr, 
and Mo isotopes was studied.
Finally, large scale studies of quadrupole correlation energies
and ms charge radii, based on both Skyrme-SLy4 \cite{bender06}
and Gogny-D1S \cite{delaroche} effective interactions have been 
carried out recently.
In summary, most of the theoretical nuclear models predict increasing
deformations up to $N=60$, where oblate and prolate shapes are
developed and exhibit minima at close energies. Which of them 
becomes the ground state depends on subtle details of the calculation.
At variance with the calculations mentioned above, in the present work
we also consider odd-$A$ nuclei in the framework of the EFA.

In the next figures we can see the results obtained for 
$\delta \langle r^2_c \rangle ^{50,N}= \langle r^2_c 
\rangle ^N - \langle r^2_c \rangle ^{50}$ 
in (a), for $S_{2n}$ energies in (b), and for $S_n$ energies in (c). 
Fig. \ref{fig_sr_r2} corresponds to Sr, Fig. \ref{fig_zr_r2} to Zr, 
and Fig. \ref{fig_mo_r2} to Mo isotopes.
In the three figures, results are shown as a function of the 
neutron number for all possible values of $N$, even or odd, in the
range of interest.
Experimental data have been taken from the mass measurements 
\cite{hager06} and from laser spectroscopy experiments 
carried out at ISOLDE/CERN and at the IGISOL facility of the University 
of Jyv\"askyl\"a \cite{charlwood,buchinger,campbell}.
Focusing on the experimental data, a consistent picture emerges.
Besides the abrupt decrease of $S_{2n}$ and $S_n$ at $N=50$ 
corresponding to the shell closure, the evolution of the $S_{2n}$ 
and $S_n$ along the isotopic chains shows a change in the tendency 
around $N=60$ in Sr and Zr isotopes. This suggests a change in the 
ground-state shape of these isotopes. On the other hand, the chain of
Mo isotopes shows a smoother behavior. These observations are confirmed 
by laser spectroscopy experiments measuring the nuclear ms radius. 
In this case, the shape change in Sr and Zr is observed in the form of 
a sudden increase of the ms charge radii at $N=58-60$. Again, the Mo 
isotopic chain \cite{charlwood} displays a smooth behavior as the 
neutron number increases.

In these figures we also see the calculated values using the oblate 
(down triangles), prolate (up triangles) and spherical (dots) shapes 
corresponding to the local minima in the PECs.
We also plot by open circles the theoretical values that correspond to 
the ground states of the corresponding isotopes.
In general, the measured $S_{2n}$ are reproduced reasonably well. The 
calculated shell gap at $N=50$ is larger than observed, but this is a 
well known feature related to any mean-field calculation. The 
discrepancy can be reduced considering dynamical correlations beyond 
mean field \cite{bender06,delaroche}. 
Between $N=52$ and $N=58$, the $S_{2n}$ energies are underestimated by 
the calculations, while they are much better reproduced beyond $N=60$. 
In our calculations a change in the tendency is observed at $N\sim 60$, 
which is more pronounced in Sr and Zr isotopes and almost disappears 
in Mo isotopes. However, the shoulder is less apparent than the 
experimental one due to the underestimation of the data below $N=60$.
Notice that the open circles at $N=59$ in $S_{2n}$ and $S_n$ do not
coincide with any specific deformation (oblate or prolate) because they 
correspond to the difference between the prolate $N=59$ and the oblate 
$N=58$ in $S_n$ and $N=57$ in $S_{2n}$.
In the case of $S_n$ energies, the amplitude of the odd-even
staggering in $S_n$ is well reproduced by the calculations indicating
the validity of our theoretical description of odd-A nuclei.
In general, the agreement is fairly good below $N=50$. Then, the 
calculations underestimate the measured $S_n$ values between $N=50$ and 
$N=60$, being the net effect a displacement to slightly lower energies. 
The agreement improves substantially for heavier isotopes.

The evolution of the nuclear charge radii can be seen in panels (a) of 
Figs. \ref{fig_sr_r2}, \ref{fig_zr_r2} and \ref{fig_mo_r2}, 
where the radius, relative to that of the $N=50$ isotope, 
is plotted as a function of neutron number. 
Results for both even-even and odd-A isotopes are included in
the plot. It is worth pointing out the almost negligible odd-even
staggering observed both in the theoretical predictions and
the experimental data.
In Fig.  \ref{fig_sr_r2}, for Sr isotopes, we can see that the 
calculations follow nicely the measurements. There is a smooth 
increase of $\delta \langle r^2_c \rangle$ up to $N=58$, then a sudden 
jump occurs at $N=60$ and again the increase is smooth for heavier 
isotopes. The encircled symbols indicate the shapes corresponding to 
the ground states obtained in our calculations. 
We can see that the lighter isotopes are spherical, 
then they change into oblate very smoothly and at about $N=60$ they 
become prolate. The observed jump corresponds to the transition from 
the oblate to the prolate shape since the oblate deformation is placed 
at $\beta \sim -0.2$ while the prolate one appears with a different
magnitude ($\beta \sim 0.4$). Fig. \ref{fig_zr_r2}(a) shows similar 
results for Zr isotopes. Here, the spherical shapes account for the 
behavior of $\delta \langle r^2_c \rangle$ up to $N=56$. From $N=56$ 
up to $N=60$ there is a smooth transition to oblate shapes that 
become the ground states and reproduce quite well the experiment. 
Above $N=60$ we obtain prolate ground states with radii in agreement
with the observed jump. For heavier isotopes (beyond $N=66$) we 
obtain again oblate shapes but there is no information in this region.
Finally, Fig. \ref{fig_mo_r2}(a) for Mo isotopes shows that the lightest 
isotopes are spherical changing into oblate shapes and increasing the   
$\delta \langle r^2_c \rangle$ very smoothly. 

In general, we observe that the calculations from $N=54-60$ with 
spherical ground states underestimate the data for the nuclear radius
in the three isotopic chains. 
One should notice that these spherical solutions
are not very sharp but shallow minima as $N$ 
increases. Thus, a possible explanation for the discrepancy in
$\delta \langle r^2_c \rangle$ could be that configuration mixing
plays an important role in these isotopes and the actual ground 
state will have contributions not only from the spherical 
configuration, but also from neighbor deformed states that will
increase slightly the charge radii.
In the case of Sr isotopes, the sudden change in $\delta \langle r^2_c \rangle$
occurs experimentally between $N=59$ and $N=60$, whereas theoretically it
appears between $N=58$ and $N=59$. Similarly in Zr isotopes we obtain the 
change between $N=60-61$, while experimentally is observed between $N=59-60$.
We do not think this discrepancy is significant since it is related to 
the subtle competition between prolate and oblate shapes. We should 
notice that in the isotopes where the shape is changing we get practically 
degenerate energies for oblate and prolate deformations and thus, tiny changes 
in the details of the calculations can lead to a different ground state. 
In particular, if the ground state in $N=59$ in Sr were oblate instead of 
prolate (they are separated by less than 0.5 MeV in the calculations)
we will get agreement with the experiment. Similarly, in Zr isotopes, if 
$N=60$ were prolate instead of oblate (separated by 0.2 MeV in the calculations)
we would match the experimental jump.
In Fig. \ref{fig_mo_r2}(a) we also recognize a difficulty in the 
reproduction of the data in the heavier isotopes, since the oblate 
shapes, which are the ground states, underestimate them.  However, 
we can see that the prolate shapes, which are close in energy, agree 
with the data. To get a further insight into the reason for this 
discrepancy and for the fact that the observed jump at $N=60$ in Sr 
and Zr almost vanishes in Mo, we have performed triaxial calculations
for the critical isotopes around $N=60$.
Similar calculations can also be found in Ref. \cite{webpage}.
We can see in Figs. \ref{fig_sr_tri}, \ref{fig_zr_tri} and 
\ref{fig_mo_tri} the $Q-\gamma$ plots for $N=$58, 60 and 62 in Sr 
(Fig. \ref{fig_sr_tri}) and Zr (Fig. \ref{fig_zr_tri}), as well as 
for $N=$58, 60, 62, 64, 66 and 68 in Mo isotopes (Fig. \ref{fig_mo_tri}). 
In these figures we can see that in Sr and Zr isotopes the transition 
from oblate to prolate at $N=60$ is manifest, suddenly changing from 
deformations with $Q\sim 2$ b in the oblate sector to $Q\sim 5$ b in 
the prolate one. On the contrary, in Fig. \ref{fig_mo_tri} for Mo 
isotopes we see that the oblate shape at $N=58$ becomes gradually 
triaxial as $N$ increases. An island of triaxiality is apparent from 
$N=60$ up to $N=68$. We have calculated the charge radii corresponding 
to these triaxial configurations that become ground states
and have added them in Fig. \ref{fig_mo_r2}(a) with open squares. 
The new $\delta \langle r^2_c \rangle$ values for these isotopes are now
very close to the axial-prolate values and agree very nicely with the 
experiment. One should notice that one important ingredient for
the agreement achieved is that the location of the quadrupole $Q$-value
for the triaxial minima is much closer to the axial prolate minima
than to the oblate ones, which are lower.

To further illustrate the  emergence and competition of deformed 
configurations in this mass region, we show in Fig. \ref{fig_zr_sp},
for the case of $^{100}$Zr ($Z=$40, $N=$60), the proton and neutron 
single-particle energies (SPE) as functions of the axial quadrupole 
moment $Q_{20}$. Fermi levels are plotted with thick dotted lines.
These diagrams help us to identify the regions of low level density, 
which favor the onset of deformation (Jahn-Teller effect), 
as well as to stress the important role of the interplay between the 
proton $\pi g_{9/2}$ and the neutron $\nu h_{11/2}$ orbitals 
(Federman-Pittel effect) to generate deformed configurations \cite{ours}.
In the plot for the proton SPE we observe an energy gap below the 
Fermi level at  $Q_{20}=5$ b, which favors the onset of prolate 
deformation in Sr isotopes. On the other hand, above the Fermi energy,
the low level density on the oblate sector favors 
oblate configurations in Mo isotopes. In the case of neutrons the 
energy gap below the Fermi level at $Q_{20}=-2.5$ b favors oblate 
shapes in lighter isotopes ($N<60$), whereas the energy gap above the 
Fermi level at $Q_{20}=5$ b favors prolate shapes in heavier isotopes 
($N>60$). Thus, these simple ideas offer a qualitative understanding 
of the various mechanisms leading to deformation in this mass region.

\section{Conclusions}

We have used  self-consistent HFB calculations based on the
interaction Gogny-D1S to study neutron separation energies and charge 
radii in neutron-rich Sr, Zr, and Mo isotopic chains. Our primary aim 
has been to search for signatures of structural changes, and more 
specifically, for shape transitions in these observables. We have 
found these correlations and specifically a remarkable connection 
between shape transitions and sudden changes in the behavior 
of the isotopic dependence of the nuclear charge radii. The different 
sensitivities of $S_{2n}$ and  $\delta \langle r^2_c \rangle$
can be understood because shape transitions take place in this mass 
region through isotopes characterized by shape coexistence, where 
the energies of the various shapes are almost degenerate. 
Thus, neutron separation energies are not particularly
sensitive to these changes. On the other hand, the sensitivity to shape 
transitions is enhanced for charge radii, specially when the transition
takes place suddenly between nuclear shapes at different absolute values
of the deformation parameter.

As compared to Sr and Zr isotopes, we have found that Mo isotopes 
exhibit a smoother increase in the charge radii with the number
of neutrons, which is in good agreement with the experimental data.
Triaxial degrees of freedom are required to get this agreement
beyond $N=60$. From triaxial calculations we have shown that
Sr and Zr isotopes suffer a sharp transition from oblate to prolate
shapes at $N=60$. On the other hand Mo isotopes display a
smooth transition through a wide region of triaxiality.

In summary, we have shown the ability of the isotopic differences in 
nuclear charge radii to signal structural changes related to deformation. 
We have also demonstrated the capability of HFB Gogny-D1S calculations 
to reproduce those features and therefore to make predictions in
unexplored regions.

\noindent {\bf Acknowledgments}

This work was supported by MICINN (Spain) under research grants 
FIS2008--01301, FPA2009-08958, and FIS2009-07277, as well as by 
Consolider-Ingenio 2010 Programs CPAN CSD2007-00042 and MULTIDARK 
CSD2009-00064. One of us (R.R.) would like to thank both Prof. 
J. \"Aysto and Dr. I. Moore from the University of Jyv\"askyl\"a 
for valuable discussions.

\end{document}